\documentclass{article}

\usepackage[english]{babel}

\usepackage[letterpaper,top=2cm,bottom=2cm,left=3cm,right=3cm,marginparwidth=1.75cm]{geometry}

\usepackage{authblk}
\usepackage{amsmath}
\usepackage{graphicx}
\usepackage{soul}
\usepackage[colorlinks=true, allcolors=blue]{hyperref}
\usepackage{dsfont}
\usepackage[dvipsnames]{xcolor}

\title{Time-Delayed Koopman Network-Based Model Predictive Control for the FRIB RFQ}
\author{Jinyu Wan\thanks{Corresponding author: Jinyu Wan (email: wan@frib.msu.edu)}}
\author{Shen Zhao}
\author{Wei Chang}
\author{Yue Hao\thanks{Corresponding author: Yue Hao (email: haoy@frib.msu.edu)}}
\affil{\small Facility for Rare Isotope Beams, Michigan State University, East Lansing, Michigan 48824, USA}
\begin{document}
\maketitle

\begin{abstract}
The radio frequency quadrupole (RFQ) at the Facility for Rare Isotope Beams (FRIB) is a critical device for accelerating heavy-ion beams from 12 keV/u to 0.5 MeV/u for state-of-the-art nuclear physics experiments. Efficient control of the RFQ resonance frequency still remains a challenge because the temperature-sensitive frequency is solely control by a cooling water system, exhibiting complicated transport delay and non-linearity in the heat transfer processes. In this work, we propose a long-short-term memory (LSTM)-based Koopman network model that can learn the time-delayed correlations hidden in historical operating data. It is proven that the model can effectively predict the behavior of the RFQ resonance frequency using historical data as inputs. With this model, a model predictive control (MPC) framework based on the Newton-Raphson method is proposed and tested. We demonstrate that the MPC framework utilizing the deep learning model is able to provide precise and rapid control for the RFQ frequency detuning, reducing the control time by half compared to the proportional-integral-derivative (PID) controller.
\end{abstract}

\section{Introduction}
The Facility for Rare Isotope Beams (FRIB) is a world-leading scientific user facility for nuclear physics research of rare isotopes, which started serving a global user community in May, 2022 \cite{1}. The radio-frequency quadrupole (RFQ) at the FRIB is a 4-vane type cavity designed to accelerate heavy ion beams from 12 keV/u to 0.5 MeV/u for state-of-the-art nuclear physics experiments \cite{2}. Due to the RF heating, thermal-induced frequency detuning can significantly shift the RFQ resonance frequency, diminishing quality of the accelerated beam \cite{3}. The capacity to mitigate and control this frequency detuning is paramount to ensure the reliable operation of the entire accelerator system. However, the frequency is solely controlled by a cooling water system, giving rise to a complicated control challenge exacerbated by water transport delay and heat transfer time. Currently, proportional–integral–derivative (PID) controller is adopted in RFQ frequency control, which usually needs tens of minutes to minimize the frequency detuning. This is particularly evident during the RFQ startup phase \cite{4}. The lengthy control time required by the PID controller serves as a compelling motivation for the application of model predictive control (MPC) \cite{5}.

MPC is a widely-used control strategy for optimizing the operation of dynamical systems. In MPC, control actions are usually chosen by optimizing a specific objective function obtained with an accurate predictive model, thereby offering the potential for faster and smoother control. To apply MPC to the RFQ frequency control, one of the foremost challenges is to construct a fast and accurate model. Due to the indirect and time-delayed nature of the water cooling control, coupled with the complicated variations in thermal expansion and contraction rates throughout the cavity geometry of the RFQ, it is exceptionally challenging to construct a comprehensive physics model for accurately predicting RFQ frequency variations based on control parameters \cite{6}. In recent years, deep learning approaches have garnered significant attention across various domains of physics research, such as quantum physics, astrophysics, condensed matter physics, nuclear physics and accelerator physics \cite{7,8,9,10,11,Edelen16, Edelen16-1}. Particularly, the application of deep learning-based Koopman operator that linearizes intricate nonlinear dynamics, has emerged as as a leading candidate for predicting nonlinear dynamical systems \cite{14,15,16}.

The Koopman operator theory is a fundamental concept in the field of nonlinear dynamical systems with a rich history dating back to 1930s \cite{17}. The basic idea of the Koopman operator theory is providing alternative descriptions of dynamical systems, namely, observables, which can be advanced by an infinite-dimensional linear operator known as the Koopman operator. The predictable linearized systems obtained through the Koopman analysis facilitate straightforward estimation and control of complicated nonlinear dynamical systems \cite{18}. The eigenvalues and eigenfunctions of the Koopman operator can contribute to the comprehension of important properties underlying nonlinear systems, such as ergodic and periodic partitions \cite{19,20,21}. 

The linearization of a nonlinear system usually yields an infinite dimension linear system. It is still necessary to make finite-dimensional approximations in practical applications. Numerical methods, such as Dynamic Mode Decomposition (DMD) \cite{22}, extended DMD (eDMD) \cite{23}, and variational approach for conformation dynamics (VAC) \cite{24}, are often used to approximate the Koopman operator. However, these methods can be computationally intensive. In recent years, there has been a growing trend in utilizing machine learning techniques to efficiently discover representations of the Koopman operator from data. For example, in \cite{14}, a modified auto-encoder was used to linearize nonlinear systems, embedding the dynamics on a low-dimensional manifold, and in \cite{15}, a deep neural network was utilized to represent the Koopman operator, demonstrating its efficacy in reinforcement learning-based control. However, conventional Koopman operator and related deep learning-based variants are primarily applied to autonomous dynamical systems that evolve over time without external intervention or influence. Although some studies also explore controlled systems with external control inputs that can influence the system's behavior, as demonstrated in \cite{15}, the control inputs are often assumed to have instantaneous impacts on the system, which may not be suitable for addressing the dynamics of a system subject to time delays in its control responses.

In this work, we develop a long-short-term memory (LSTM)-based Koopman network \cite{25}. The LSTM network, known as one of the most potent recurrent neural networks (RNNs) \cite{26}, is capable of embedding time-series data into a scalar representation by capturing essential temporal correlations and patterns hidden in the data \cite{Edelen20}. Unlike the auto-encoder structure in \cite{14}, we employ a similar technique in eDMD where the system behavior becomes a component of the observables for the Koopman operator. This structure avoids using an additional encoder to reproduce the prediction, which is particularly challenging in time series generation. The model is trained with historical operating data from the FRIB RFQ and tested with data collected after the training phase. We demonstrate the effectiveness of the model to accurately and rapidly predict the variance of the RFQ resonance frequency over time up to 300 seconds. With this trained model, we employ a Newton-Raphson \cite{28} optimizer for the implementation of MPC to control the frequency of the RFQ, optimizing the frequency detuning to zero by adjusting two valves controlling the cooling water system. Compared to the PID controller, MPC can reduce the control time by approximately a factor of 2.

\section{Frequency control of the FRIB RFQ}
\begin{figure}
    \centering
    \includegraphics[width=0.55\linewidth]{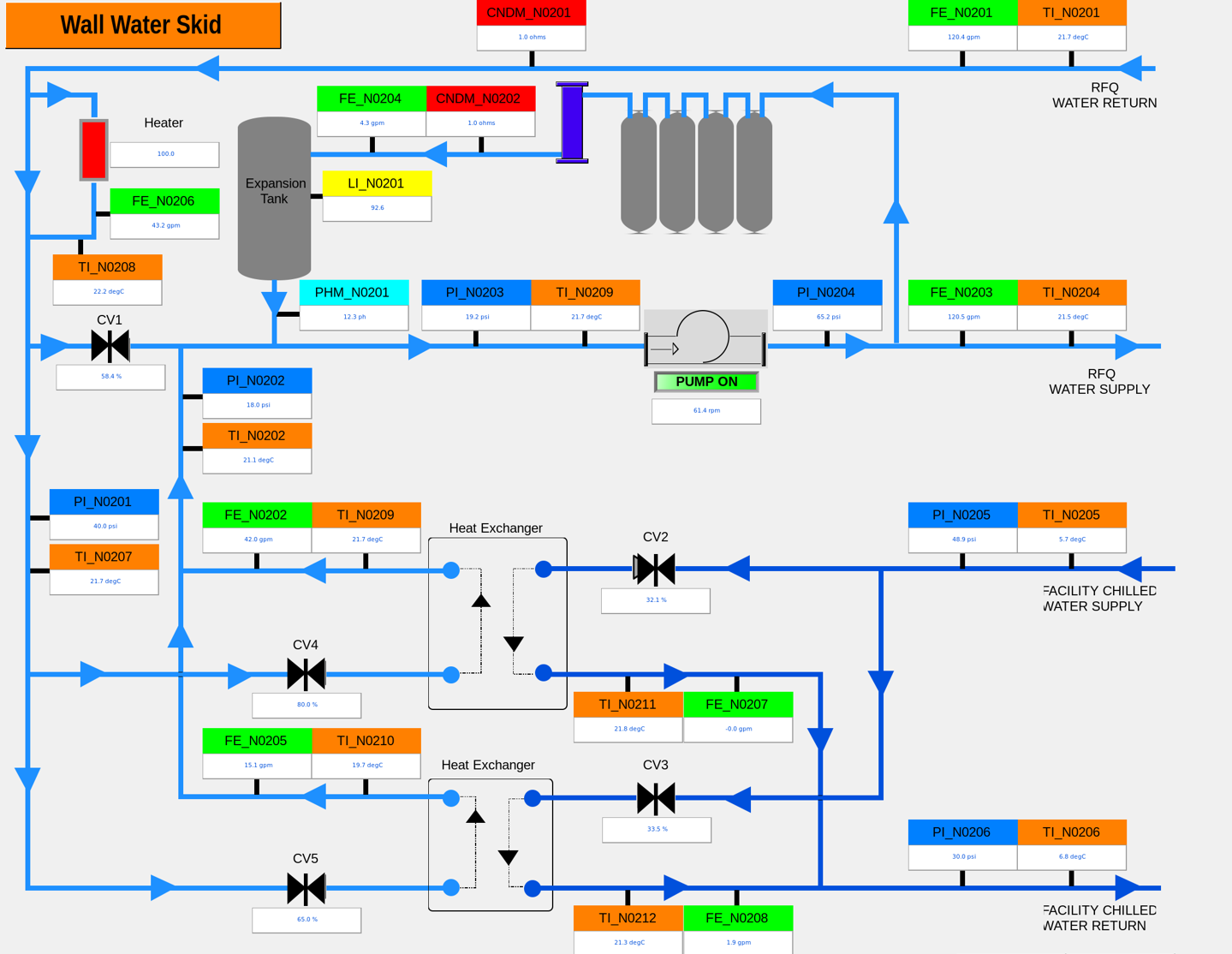}
    \caption{Schematic illustration of the water skid for wall temperature control. The variables, e.g., TIN0101, FEN0101 and PIN0101, represent water temperature, water flow, and pressure measured at specific locations, respectively. CV1 to CV5 are valves controlling the water flow.}
    \label{fig:1}
\end{figure}
The FRIB RFQ comprises five longitudinal segments, each approximately one meter in length. Twenty-seven fixed slug tuners are positioned along the cavity to control the field profile and the cavity resonance frequency. The slug tuners are cut and fixed by a tuning algorithm to tune the RFQ resonance frequency to the target value, 80.5 MHz, with a low RF power \cite{2}. After the slug tuners are fixed, the cooling water system becomes the sole mean for controlling the resonance frequency. Fig. 1 illustrates the closed-loop water skid used to control the wall temperature. Similar to the wall skid, a vane skid is also designed. The two separate skids allow for independent control for the wall and vane temperature.

As seen in Fig. 1, the cooling water is sourced from the chilled water supply and is delivered to the supply end of the RFQ. Then the used cooling water is collected at the return end of the RFQ, where it can be recycled. Five valves, specifically, CV1 to CV5, are employed to control the flow of the cooling water. Currently, only CV2 and CV3 are in active use for control, among which CV2 is more sensitive than CV3. In practical control process, CV3 is usually adjusted first, providing fine-tuning of the water flow. If the setting of CV3 reaches the upper or lower bound, CV2 will be tuned as coarse adjustment. Due to the nonlinear behavior of the valves and the complicated heat transfer process \cite{morris2018rf}, the frequency control can be challenging in practice, which usually takes tens of minutes with a PID controller.

The operation of the cooling water system involves over 150 relevant variables in the system. The substantial number of variables presents a challenge to use all of them for constructing an effective deep learning model. To address this issue, based on expert's input, we have ultimately selected a subset of variables considered most crucial to train our deep learning model. These chosen variables are listed in Table 1. Besides the variables of the water cooling system, the forward power, reflected power and electric field in the RFQ cavity are also included.

\begin{table}
    \centering
    \caption{Selected variables for deep learning model training.}
\begin{tabular}{cc}
    \hline
    Variable name & Description\\
    \hline
    $FE\_LCW1:TI\_N0104:T\_RD$ & Vane temperature Pos. 4\\
    $FE\_LCW1:FE\_N0103:F\_RD$ & Vane flow\\
    $FE\_LCW2:TI\_N0205:T\_RD$ & Wall temperature Pos. 5\\
    $FE\_LCW2:TI\_N0207:T\_RD$ & Wall temperature Pos. 7\\
    $FE\_LCW2:PI\_N0201:P\_RD$ & Wall pressure Pos. 1\\
    $FE\_LCW2:PI\_N0205:P\_RD$ & Wall pressure Pos. 5\\
    $FE\_LCW1:CV\_N0102:POS\_RD\_PLC$ & $CV2$ of vane skid\\
    $FE\_LCW1:CV\_N0103:POS\_RD\_PLC$ & $CV3$ of vane skid\\
    $FE\_LCW2:CV\_N0202:POS\_RD\_PLC$ & $CV2$ of wall skid\\
    $FE\_LCW2:CV\_N0203:POS\_RD\_PLC$ & $CV3$ of wall skid\\
    $FE\_RFQ:RFC\_D1005:E\_RD\_CAVS$ & Electric field\\
    $FE\_RFQ:RFC\_D1005:POWR\_RD\_FWDS$ & Forward power\\
    $FE\_RFQ:RFC\_D1005:POWR\_RD\_RFLS$ & Reflected power\\
    $FE\_RFQ:RFC\_D1005:FR\_RD\_ERR$ & Frequency detuning\\
    \hline
\end{tabular}
\label{tab:table1}
\end{table}

\section{Method}
\subsection{Koopman operator theory}
Consider a discrete autonomous nonlinear dynamical system,
\begin{equation}
x_{t+1} = f(x_t)
\end{equation}
\begin{equation}
\phi_t = \phi(x_t)
\end{equation}
where $x_t \in \mathds{R}^n$ represents the system behavior at time $t$, $f$ is a continuously differentiable nonlinear function, and $\phi_t \in \mathds{R}^m$ represents a possible observable of $x_t$. Koopman operator theory provides an infinite-dimensional linear operator $\mathcal{K}$ (Koopman operator) that can linearize the evolution of the observable $\phi_t$,

\begin{equation}
\phi_{t+1} = \mathcal{K}\phi_t.
\end{equation}

For controlled systems with external intervention or influence affecting the system state, it is common to extend the Koopman operator as
\begin{equation}
\phi(x_{t+1},u_{t+1}) = \mathcal{K}\phi(x_t,u_t)
\end{equation}
where $u_t$ represents the external control vector at time $t$. 

The finite-dimensional representation of the Koopman operator is often approximated by DMD and eDMD \cite{22,23}. It is worth noting that in the eDMD, the observables $\phi$ are often represented as polynomials of the systems state $x$, which allows for the direct extraction of $x$ at its first order.

\subsection{LSTM-based Koopman network model}
\begin{figure}
    \centering
    \includegraphics[width=0.5\linewidth]{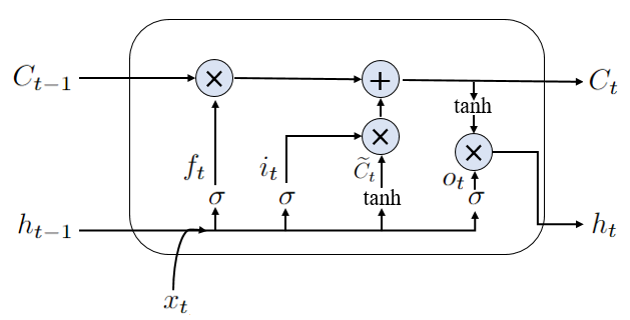}
    \caption{Schematic diagram of an LSTM cell. $x_t$ is the network input at $t$th time step. $f_t$, $i_t$ and $o_t$ represent the forget gate, input gate and the output gate, respectively. $\sigma$ and tanh represent activation functions. $h_t$ is the hidden state of the cell. $C$ represents the cell state and $\widetilde{C}$ is the updated cell state.}
    \label{fig:Fig. 2}
\end{figure}
\begin{figure}
    \centering
    \includegraphics[width=0.9\linewidth]{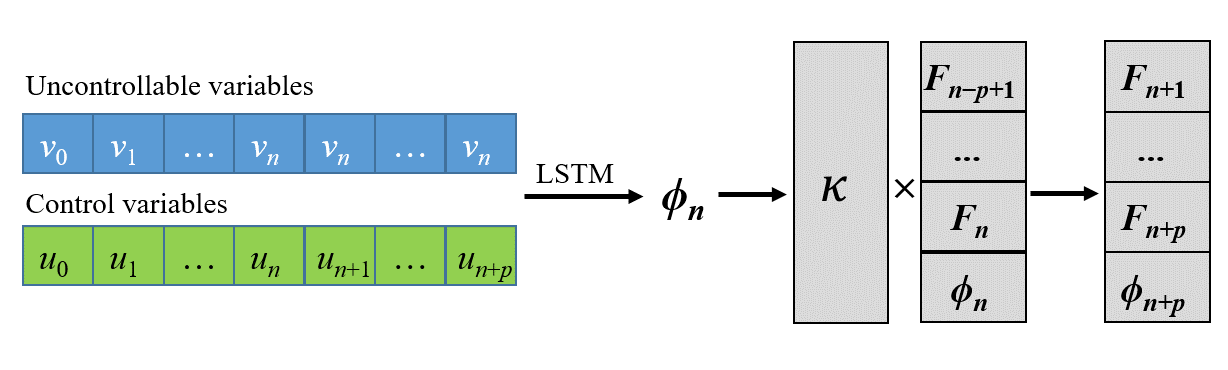}
    \caption{Schematic representation of the LSTM-based Koopman network model. $\{v_0, v_1, ..., v_n\}$ and $\{u_0, u_1, ..., u_n\}$ represent time-series data of the environment variables and control actions, respectively. These time-series variables have time-delayed impacts on the final output. $\phi_n$ represents a observable function. $\mathcal{K}$ represent the Koopman network that advances $\{F_{n-p+1}, ..., F_n, \phi_n\}$, where $F_n$ is frequency detuning at time $n$.}
    \label{fig:3}
\end{figure}
In this work, our data-driven model is to predict the RFQ frequency detuning starting from time step $n+1$ to $n+p$ for the control purpose. To address the control problems subject to time delays in its control responses, a LSTM network is used to embed the time-series data into a scalar representation. LSTM is a powerful variant of RNNs that overcomes stability challenges encountered in traditional RNNs, such as vanishing gradient, which makes it a practical choice for wide applications\cite{25}. A typical LSTM cell consists of three gates controlling the information flow through the network, namely, input gate, forget gate and output gate (see in Fig. 2). The three gates allow the LSTM to selectively add or forget information, and let information pass through to the next cell, respectively. The architecture of an LSTM cell can be described as
\begin{equation}
f_t = \sigma(W_f\cdot[h_{t-1},x_t]+b_f)
\end{equation}
\begin{equation}
i_t = \sigma(W_i\cdot[h_{t-1},x_t]+b_i)
\end{equation}
\begin{equation}
\widetilde{C}_t = \operatorname{tanh}(W_C\cdot[h_{t-1},x_t]+b_C)
\end{equation}
\begin{equation}
C_t = f_tC_{t-1} + i_t\widetilde{C}_t
\end{equation}
\begin{equation}
o_t = \sigma(W_o\cdot[h_{t-1},x_t]+b_o)
\end{equation}
\begin{equation}
h_t = o_t\operatorname{tanh}(C_t)
\end{equation}
where $W$ and $b$ represents weights and biases for the three gates. The information is propagated inside the cell, resulting in the cell state. The states are propagated ahead through the network.

Fig. 3 shows the schematic diagram of our model. The time-series inputs of the model consist of the historical data of selected relevant variables $\{v_0, v_1, ..., v_n\}$ collected prior to the control action, including environment variables such as water temperature, water flow, and water pressure, and the frequency detuning itself. Additionally, previous control actions {$u_0, u_1, ..., u_n$}, as well as future control actions $\{u_{n+1}, u_{n+2}, ..., u_{n+p}\}$ are also included. We extend the sequence $\{v_0, v_1, ..., v_n\}$ from length $n$ to $n+p$ by duplicating these variables at the $n$th time step, ensuring that both environment variables and control actions can be used as inputs for the same LSTM network. This duplication also enforces the importance of the variables at the $n$th time step in the training. Alternatively, the extended portion can be filled with 0 if the enforcement is not needed. 

With the LSTM network, an observable function $\phi_n$ is obtained. Inspired by the polynomial observable function often utilized in eDMD, we combine $\phi_n$ with historical frequency detuning $\{F_{n-p+1}, ..., F_n\}$ to construct a new observable function $\{F_{n-p+1}, ..., F_n, \phi_n\}$, where $p$ is the size of a look-back window. A Koopman operator $\mathcal{K}$ represented by a single-layer non-bias neural network acts on the new observable function to advance the observable by $p$ time steps, resulting in the predicted frequency detuning $\{F_{n+1}, ..., F_{n+p}\}$.

\subsection{Model predictve control}
With the LSTM-based Koopman network model, we will demonstrate the implementation of MPC for the RFQ frequency control in Section 4.3. An MPC controller typically consists of a predictive model and an optimizer. The optimizer is used to optimize the objective function, i.e., frequency detuning in this scenario, by adjusting the control actions. It is important to choose an appropriate optimizer for MPC, as it directly affects the efficiency and effectiveness of the controller. In the exploration of different optimization algorithms, including least squares, genetic algorithms and the Newton-Raphson method, the Newton-Raphson method standing out for its high efficiency and dependability is ultimately chosen for optimizing the frequency detuning.

We denote the frequency detuning controlled by the control action $\mathbf{u}$ as $F(\mathbf{u})$. When the control action is changed by $\Delta \mathbf{u}$, the new value of $F(\mathbf{u}+\Delta \mathbf{u})$ can be written as
\begin{equation}
    F(\mathbf{u}+\Delta \mathbf{u}) \approx F(\mathbf{u}) + \nabla F(\mathbf{u})\Delta \mathbf{u} + \frac{1}{2} \Delta \mathbf{u}^T H(\mathbf{u})\Delta \mathbf{u}
\end{equation}
where $\nabla F(\mathbf{u})$ is the gradient of $F$ at $\mathbf{u}$, and $H(\mathbf{u})$ is the Hessian matrix of $F$ at $\mathbf{u}$. To find a local minimum of $F(\mathbf{u})$, we can solve the following equation to make the gradient of Eq. (11) to be zero.
\begin{equation}
    \nabla F(\mathbf{u}) + H(\mathbf{u})\Delta \mathbf{u} = 0
\end{equation}
Thus we can obtain
\begin{equation}
    \Delta \mathbf{u} = -H(\mathbf{u})^{-1}\nabla F(\mathbf{u})
\end{equation}

With the predictive model, we can easily obtain the approximated gradient and Hessian matrix with the following numerical method,
\begin{equation}
    \nabla f(\mathbf{u}) \approx \left[ \frac{f(\mathbf{u} + \epsilon \mathbf{e}_1) - f(\mathbf{u} - \epsilon \mathbf{e}_1)}{2\epsilon}, \ldots, \frac{f(\mathbf{u} + \epsilon \mathbf{e}_n) - f(\mathbf{u} - \epsilon \mathbf{e}_n)}{2\epsilon} \right]^T
\end{equation}
\begin{equation}
    H_{ij}(\mathbf{u}) \approx \frac{f(\mathbf{u} + \epsilon(\mathbf{e}_i + \mathbf{e}_j)) - f(\mathbf{u} + \epsilon\mathbf{e}_i) - f(\mathbf{u} + \epsilon\mathbf{e}_j) + f(\mathbf{u})}{\epsilon^2}
\end{equation}
where $\mathbf{u} = [u_1, u_2, \ldots, u_n]^T$ is the control action, $\epsilon$ is a small scalar perturbation, and $\mathbf{e}_i$ is the unit vector in the direction of the $i$-th control action. Utilizing the numerically derived gradient and Hessian, we can calculate the parameter updates $\Delta \mathbf{u}$ for optimizing frequency detuning by employing Equation (13). 

\section{Results}
\subsection{Data preparation and model training}
To construct the LSTM-based Koopman model for predicting RFQ frequency detuning, we have selected 13 relevant variables as listed in Table 1, as well as the frequency detuning itself, as input variables. Variables  associated with the cooling water system, such as water temperature, pressure and flow, are considered time-delayed variables. The open voltages of flow control valves, namely CV2 and CV3 for the wall skid, are adjustable control variables. Alongside these, we include non-delayed variables, i.e., the forward power, reflected power, and electric field of the RFQ cavity.  

We have collected operating data on the selected variables for two and a half months at FRIB. Through an interpolation algorithm, all the variables are synchronized to the same timestamps with a time interval of five seconds. Based on various tests, we ultimately set the duration of input for our model to be 300 seconds, and the predicted time horizon is also 300 seconds. The data we collected is split into two parts. To evaluate the model's performance for predicting future frequency detuning, the data order remains unshuffled. The first 80\% of the data is used for training and the remaining 20\% is used for validation and testing purposes.

As shown in Fig. 3, an LSTM network and a Koopman network are constructed with Keras \cite{30}. The observable function is advanced by the Koopman network and thus the frequency detuning is predicted ahead of time. A critical parameter for this model is the dimension of the observable function. An adequately high observable dimension is important for the model to linearize this complicated nonlinear dynamical system. However, a high observable dimension can slow down the computation through the network, resulting in an undesirably slow prediction for MPC. In this work, we set the observable dimension as 64, which strikes a balance between good accuracy and fast prediction time. The number of cells in the LSTM layer. The loss function is the mean squared error between the prediction and the ground truth. The Adam \cite{31} optimizer is used to train the model with a training rate of 0.001 and a training batch size of 512. The training process takes nearly an hour on a GeForce RTX 4070 Graphics Card. The prediction for a single sample takes \textasciitilde0.04 seconds on the same device.

\subsection{Model evaluation}
\begin{figure}
    \centering
    \includegraphics[width=0.95\linewidth]{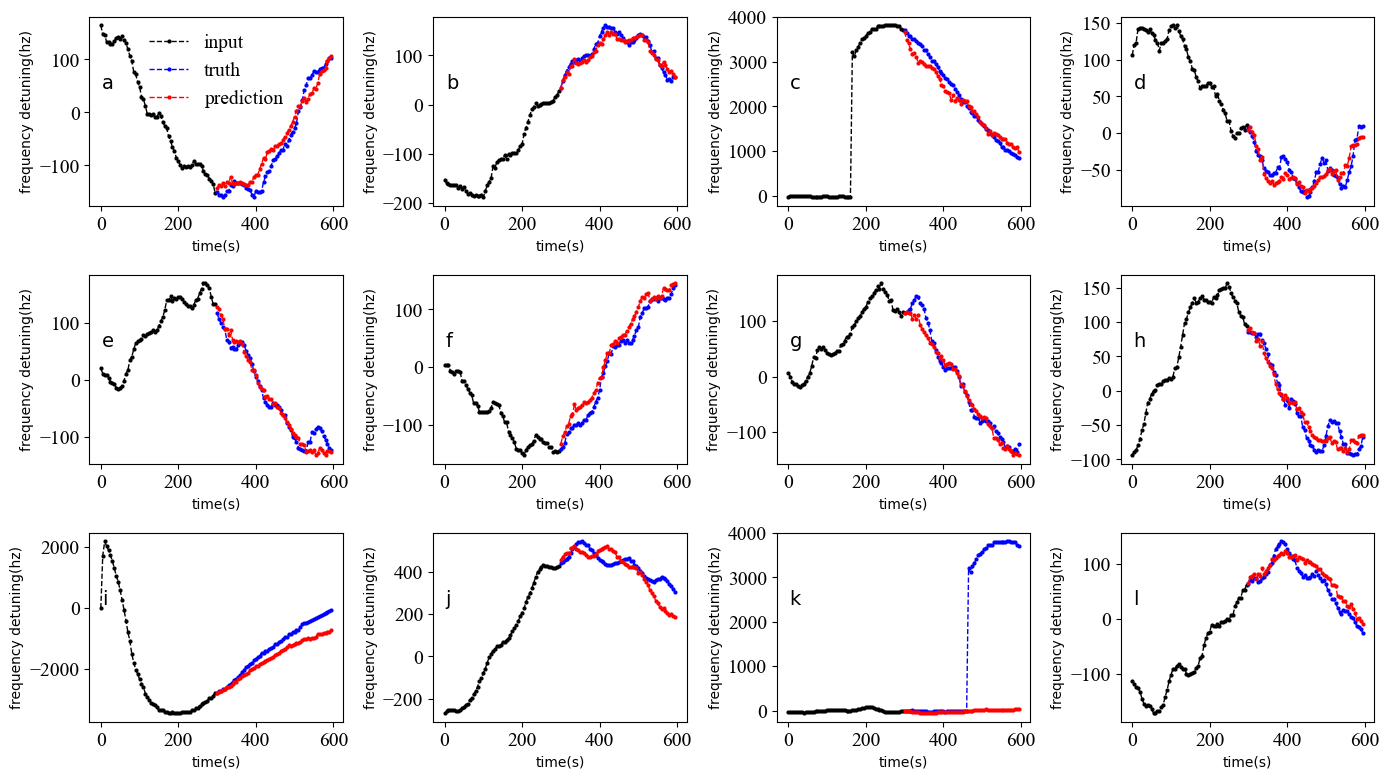}
    \caption{Frequency detuning for selected testing samples. The black dots within 0-300s represent the historical frequency of detuning. The blue and red dots within 300-600s represent the measured frequency detuning and the predicted frequency detuning, respectively.}
    \label{fig:4}
\end{figure}
\begin{figure}
    \centering
    \includegraphics[width=0.7\linewidth]{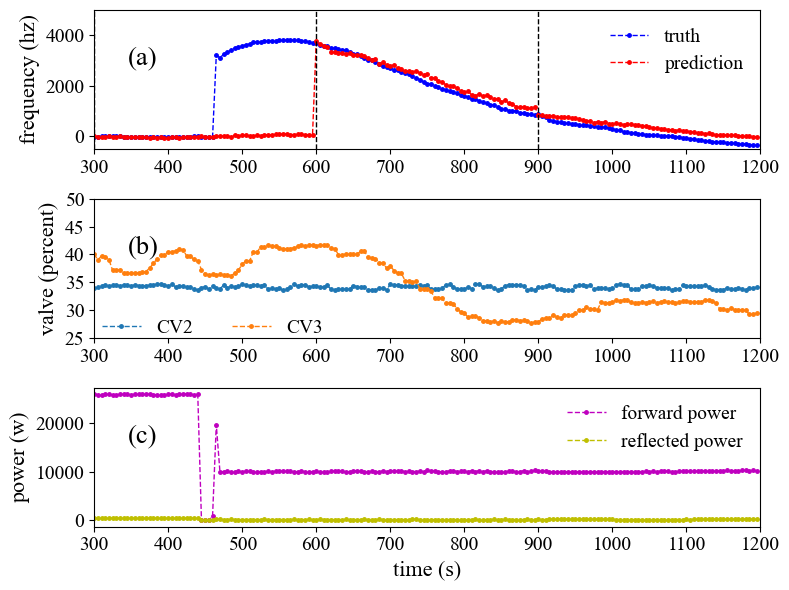}
    \caption{Detailed comparison for the testing sample in Fig. 4(k). (a) shows the variations of frequency detuning. Note that the predicted frequency detuning is split into three part, where each part represents a single prediction obtained with the LSTM-based Koopman network model. (b) and (c) show the changes in the valve settings and the cavity power over time, respectively.}
    \label{fig:5}
\end{figure}
To demonstrate our LSTM-based Koopman network model, Fig. 4 shows the model predictions for some selected testing samples. It is found that the model prediction remains reliable not only when the operation is stable with low-frequency detuning, e.g., Fig. 4(a, b, d, e, f, g, h, l), but also works effectively in case with significant frequency detuning, e.g., Fig. 4(c, i, j). The results indicate that the model effectively learns the latent time-series correlations hidden in the data, enabling accurate predictions of frequency-detuning trends for most test data across extended time horizons of up to 300 seconds. This holds true, especially, for the significantly detuning cases, which is of particular concern in the RFQ operation. 

However, it is still observed that there are a few cases where the prediction deviates significantly from the historical data, e.g., Fig. 4(k). To further investigate the reason for these false predictions, we locate the false prediction observed in Fig. 4(k) and present the changes of variables in the preceding and subsequent time period in Fig. 5. Since the environment variables and the electric field remain stable at this time, only the changes of the settings of CV2 and CV3, and the changes of forward power and reflected power are presented. 

Fig. 5 shows that a quick change of forward power occurs at \textasciitilde450s, causing a significant frequency detuning in a short period of time. The model prediction becomes unreliable after the quick change happens, exhibiting a large divergence from the measured frequency. It is reasonable because while predicting the frequency detuning within 300-600s, the input of the model is the data collected within 0-300s. Therefore, the model cannot anticipate such a quick change of forward power occurring after 300s and will not give a relevant response in the prediction. This issue can be a limitation for models with mixed time-delayed and non-delayed variables. Fortunately, forward power is not a variable that changes often. The changes in forward power often result from changes in setting point or cavity trips followed by quick recovery, which usually happens in a short period of time. As a result, the RFQ forward power is stable throughout most instances. Results in Fig. 5 indicate the prediction becomes reliable when the forward power becomes stable again.

\subsection{Demonstration of model predictive control}
To demonstrate the implementation of MPC, we select the significantly detuning sample in Fig. 4(k) as an example to control. Learned from Fig. 5, the model prediction is reliable after 600s. Thus we start the control at 600s. The optimization variables are time-series control actions of CV2 and CV3 within 600-900s, and the optimization objective is to minimize the frequency detuning within 600-900s, i.e., $\operatorname{min}\sum_{t=300}^{600}W_tF_t^2$, where $F_t$ represents frequency detuning at time $t$ and $W$ is the weight. The changes of frequency detuning are evaluated with the LSTM-based Koopman network model.

The optimization process is subject to two key constraints. First, the valve settings must remain within a reasonable range of [0, 80]. Second, any changes in the valve settings from one time step to the subsequent time step cannot exceed 1.0, which prevents abrupt and substantial changes that may damage the devices. The initial solution for the Newton-Raphson optimizer is the set points of CV2 and CV3 at 600s in Fig. 5. Considering the fact that the model become less reliable over extended periods, we choose to promptly apply the optimizer to adjust control actions, despite its capability for long-term prediction. The time interval of implementing the MPC controller is set to be five seconds, which is the same as the time interval of the model prediction. The calculation of an optimization iteration takes ~0.2s on a personal laptop, which is fast enough to ensure the implementation of online optimization with a time interval of five seconds.

\begin{figure}
    \centering
    \includegraphics[width=0.5\linewidth]{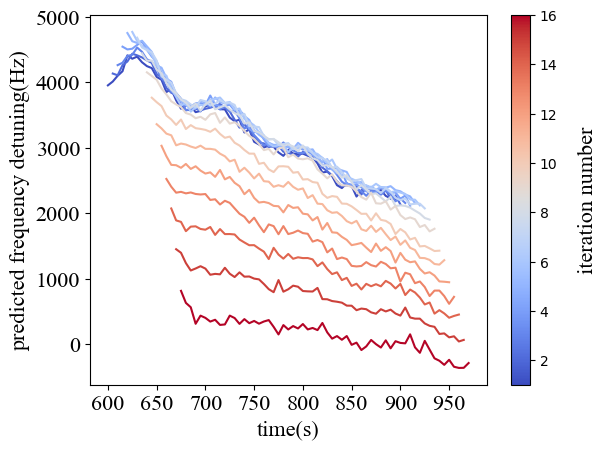}
    \caption{Optimization process of the Newton-Raphson method. The solid lines represent the predicted frequency detuning over a 300-second span. Each line originates from a distinct starting time, corresponding to a separate iteration of the optimization process. The evolution of the results indicates the application and subsequent effect of the optimized control actions determined in the preceding iterations. The color from blue to red represents the iteration from initial iterations to later ones.}
    \label{fig:6}
\end{figure}

Fig. 6 shows the evolution of the optimization. The results indicate the effectiveness of our Newton-Raphson optimizer in achieving precise control of CV2 and CV3 within only about 20 iterations, effectively eliminating the frequency detuning in  \textasciitilde350 seconds. Compared to the PID controller, the MPC controller reduces the control time by almost 50\%. A key advantage of the MPC controller is its ability to simultaneously adjust CV2 and CV3, as illustrated in Fig. 7. This concurrent manipulation exploits the complex interactions between multiple valves learned by the model to achieve more effective control. Furthermore, the MPC controller responds more swiftly to the frequency detuning due to its capacity for long-term prediction. In contrast, the PID controller must wait for a delayed response from the system, which is usually slower than MPC.

\begin{figure}
    \centering
    \includegraphics[width=0.45\linewidth]{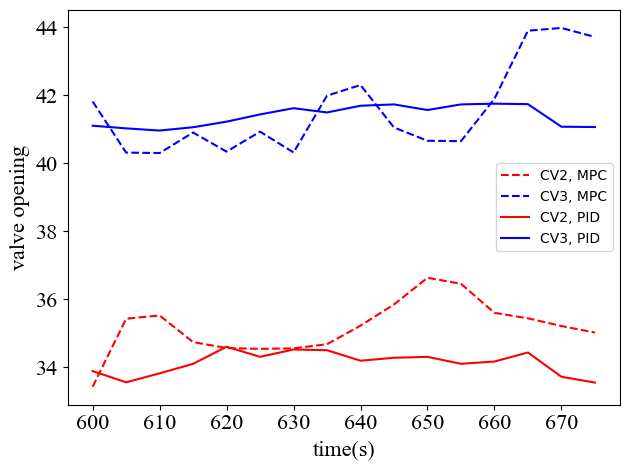}
    \caption{Comparion of control actions given by the PID controller and the MPC. The solid lines represent the control actions given by the PID controller, and the dashed lines represent the control actions given by the Newton-Raphson optimizer. Red and blue lines represent CV2 and CV3, respectively.}
    \label{fig:7}
\end{figure}
\section{Conclusion}
A novel LSTM-based Koopman network model is proposed to address the frequency detuning control problem for the FRIB RFQ. The time-series data of the RFQ system is converted into an observable function with a LSTM network. With a Koopman network, the evolution of the observable function is linearized, and the future behavior of the system states can be easily obtained without training an additional decoder. We demonstrate the model can effectively predict the behaviors of the RFQ resonance frequency up to 300 seconds within a short period of \textasciitilde0.04 seconds. 

With the well-trained RFQ frequency model, an MPC framework is proposed. The MPC is based on a Newton-Raphson optimizer that optimizes the RFQ frequency detuning by adjusting the control actions of two water flow control valves. This MPC framework, powered by the LSTM-based Koopman network, exhibits a notable advantage compared to the PID controller. With the assistance of the model's rapid prediction capability, the optimizer can simultaneously adjust the two valves that takes advantage of the intricate interplay between them. The control time for eliminating frequency detuning is reduced by half with MPC compared to using the PID controller.

The process shown in this paper provides a general framework for MPC design to handle the mixture of delayed and nondelayed control knobs. Therefore, we foresee the potential of applying the proposed method to a much broader application in complicated control problems across various scientific and industrial domains.
\section{Acknowledgement}
This work is supported by the U.S. Department of Energy Office of Science under Cooperative Agreement DE-SC0023633, the State of Michigan, and Michigan State University.


\end{document}